\documentclass[english]{PoS}
\usepackage[T1]{fontenc}
\usepackage[latin9]{inputenc}
\usepackage{units}
\usepackage{amsmath}
\usepackage{amssymb}
\usepackage{graphicx}
\usepackage[numbers]{natbib}

\makeatletter

\title{Calibration and monitoring of the ASTRI-Horn telescope by using the night-sky background measured by the photon-statistics ("variance") method}

\ShortTitle{Calibration and monitoring of the ASTRI-Horn telescope by using the night-sky background}

\author{\speaker{A. Segreto}, O. Catalano, M. C. Maccarone, T. Mineo, A. La Barbera,  for the CTA ASTRI Project\footnote{for consortium list see PoS(ICRC2019)1177}\\
		INAF - Istituto di Astrofisica Spaziale e Fisica Cosmica di Palermo, via Ugo La Malfa 153, Palermo, Italy\\
		E-mail: \email{segreto@ifc.inaf.it}}

\abstract{ASTRI-Horn is the Cherenkov telescope developed by INAF and operating in Italy on the slopes of Etna volcano. Characterized by a dual-mirror optical system and a focal plane covered by silicon photomultiplier sensors, ASTRI-Horn is a prototype of the telescopes proposed to form one of the pathfinder sub-arrays of the Cherenkov Telescope Array Observatory in Chile. The electronics of the ASTRI-Horn telescope, optimized to detect nanosecond burst of light, is not able to directly measure any continuous or slowly varying flux illuminating its camera. To measure the intensity of the night sky background (NSB) in the field of view of the telescope, the firmware of the ASTRI-Horn camera continuously performs the statistical analysis of its detector signals and periodically provides in output the "variance" of each pixel, which is linearly dependent on the rate of detected photons; in this way, an indirect, but accurate measurement of the NSB flux is obtained without interference with the normal telescope operation. In this contribution we provide an overview of several calibration and monitoring tasks that can be performed in a straightforward way by the analysis of the "variance" data such as the camera astrometry, the actual telescope orientation and the monitoring of its optical point spread function.}

\FullConference{36th International Cosmic Ray Conference -ICRC2019-\\
		July 24th - August 1st, 2019\\
		Madison, WI, U.S.A.}

\usepackage{natbib}
\date{\today}

\makeatother

\usepackage{babel}
\begin{document}

\section{Introduction}

ASTRI\nobreakdash-Horn \citep{ASTRI_Pareschi} is the end\nobreakdash-to\nobreakdash-end
prototype telescope developed by the Italian National Institute of
Astrophysics, INAF, in the context of the Cherenkov Telescope Array
Observatory, CTAO. It has been installed in Italy at the INAF station
Fracastoro (located at Serra La Nave on Mount Etna) during Fall 2014.
The ASTRI\nobreakdash-Horn telescope is characterized by a dual-mirror
optical system and a curved focal surface covered by silicon photomultiplier
(SiPM) sensors ($\unit[7\mathbin{\times}7]{mm^{2}}$) organized in
photo detection modules (PDMs) managed by a fast front-end electronics
based on the ASIC CITIROC \citep{Astri_camera}.

In a way similar to other telescopes designed to detect fast pulsed
signals (like e.g. the Fluorescence light or Cherenkov radiation produced
by air showers), ASTRI\nobreakdash-Horn adopts an AC-coupled readout
electronics that provides several advantages like, e.g., the safe
connection between electronic stages powered at different voltage
levels and keeps the working points of the signal amplifiers and the
baselines of the ADC converters stable at a predefined value, against
any variation of the external background light.

The AC\nobreakdash-coupling however, blocking any slow varying signal,
makes the telescope blind to the Night Sky Background (NSB) flux (diffuse
emission and stars) thus eliminating any direct feedback from the
sky in the telescope field of view (FoV). However, due to the fact
that the light is a sequence of photons described by a Poisson process,
the individual electronic pulses generated at the random photon arrival
times, produce fast varying fluctuations in the total electronic signal
that are not blocked by the AC\nobreakdash-coupling and, reaching
the input of the ADC converter, add a noise component which is dependent
on the rate of detected photons.

While the mean value of the ADC pedestal is therefore independent
of the NSB flux, the variance (square of standard deviation) of the
pedestal fluctuations, $\sigma^{2}$, is linearly dependent on it,
being expressed by

\[
\sigma^{2}\mathrel{=}\sigma_{\mathrm{dark}}^{2}\mathbin{+}\sigma_{\mathrm{sky}}^{2}\mathrel{=}\sigma_{\mathrm{dark}}^{2}\mathbin{+}k\mathbin{\cdot}\Phi_{\mathrm{sky}}
\]
where $\sigma_{\mathrm{dark}}^{2}$ is the intrinsic variance of the
electronics and the detector as it would be observed in perfectly
dark conditions, and $\sigma_{\mathrm{sky}}^{2}$ the variance induced
by the external NSB flux which is directly proportional to the photon
flux, $\Phi_{\mathrm{sky}}$, illuminating the detectors.

This equation shows how measurements of the pedestal variance performed
by the statistical analysis of a sequence of consecutive ADC converter
outputs, allow us to obtain an indirect measure of night sky flux
brightness without requiring any additional hardware. The statistical
accuracy that can be obtained with this method depends on the number
of ADC samples used for the analysis; denoting with $\mathord{\delta}\sigma^{2}$
the statistical uncertainty on the variance estimated from $N_{\mathrm{ADC}}$
samples, and assuming that the samples follow a normal distribution,
its fractional uncertainty is expressed by the formula \citep{Casella_Berger}

\[
\frac{\mathord{\delta}\sigma^{2}}{\sigma^{2}}\mathrel{=}\sqrt{\frac{2}{N_{\mathrm{ADC}}-1}}
\]

From this equation we obtain that by using $N_{\mathrm{ADC}}\mathrel{=2^{16}}$
samples, the variance of the pedestal noise is estimated with a statistical
uncertainty of $\mathop{\thickapprox}\unit[0.55]{\%}$ and, assuming
negligible the intrinsic dark noise, also the photon flux is estimated
with the same accuracy level.

\section{The ASTRI-Horn real time \textquotedblleft variance\textquotedblright{}
computation}

Using a method similar to the one implemented for the Fluorescence
telescopes operating at the Pierre Auger Observatory \citep{Statistical_Monitor_Kleifges,NSB_Segreto},
an algorithm implemented in the ASTRI FPGA front\nobreakdash-end
board performs in real time, for each camera pixel, the sampling of
the ADC baseline, and accumulates the sum of their values and of their
squares; the algorithm runs in parallel with the scientific data taking
discarding the samples acquired at the occurrence of camera triggers.

After a predefined number of successive ADC samples, the results of
the accumulated data are inserted in the output telemetry and from
these values, in a straightforward way, the ``variance'' maps of
the focal plane are periodically obtained. Taking into account that
the front end of the ASTRI-Horn electronics requires $\unit[16]{\mu s}$
to read-out the ADC samples of the whole detector plane, the total
time required to produce a ``variance'' map with high statistical
significance ($2^{16}$ samples) is $\mathrel{\thickapprox\unit[1]{sec}}$,
which is sufficient to monitor the temporal variations of the NSB
flux.

\begin{figure}[h]
\begin{centering}
\includegraphics[height=0.34\textwidth]{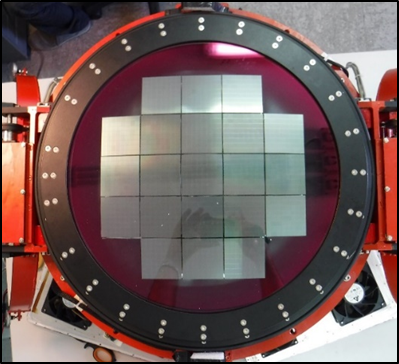}\quad{}\includegraphics[height=0.34\textwidth]{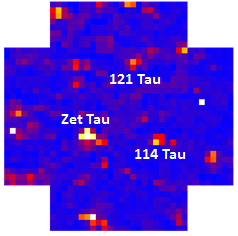}
\par\end{centering}
\caption{Left panel: the layout of the SiPM (64 pixels) tiles, each one integrated
in a PDM (Photo Detection Module), covering the ASTRI curved focal
plane. Right panel: an example of a single ``variance'' image obtained
with the telescope pointing to the Crab, where the presence of several
stars, although with a quite coarse pixel resolution, is evident.}

\label{fig:Crab_raw_frame}
\end{figure}

In Fig.~\ref{fig:Crab_raw_frame} we show an image of the ASTRI focal
plane, where the detector layout is visible, and one example of ``variance''
acquisition, obtained while the telescope was pointing in the Crab
direction, where it is well evident as several stars are clearly detected.

\section{Applications of the ``variance'' data}

There are numerous possible application of the NSB measured by means
of the \textquotedblleft variance\textquotedblright{} data: verification
of the mirror segments alignment, astrometric calibration of the camera
FoV, telescope pointing monitor, point spread function measurements,
and much more\citep{Segreto_CRIS_2018}. In the following sections,
using real data, we illustrate some of the problems that have been
solved thanks to this innovative method.

\subsection{Verification of the alignment of primary mirror segments}

\begin{figure}[h]
\begin{centering}
\includegraphics[height=0.34\textwidth]{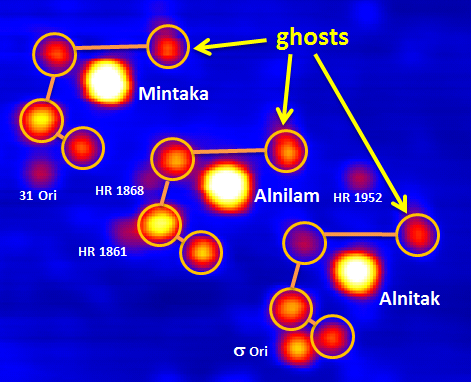}
\par\end{centering}
\caption{The first \textquotedblleft variance\textquotedblright{} sky map obtained
with the ASTRI-Horn telescope pointing to the \textquotedblleft Orion's
Belt\textquotedblright{} region; the numerous spots in repetitive
pattern around the three brightest stars are caused by several tilted
segments of the primary mirror.}

\label{Orion_belt_ghosts}
\end{figure}

The first useful application obtained by the ``variance'' sky maps
was the discovery of the misalignment of several segments of the ASTRI-Horn
primary mirror. In Fig.~\ref{Orion_belt_ghosts}, we show the sky
map that was obtained with the telescope pointing to the \textquotedblleft Orion's
Belt\textquotedblright{} where the repetitive pattern of bright spots
near the three brightest stars are a clear indication that several
segments of the ASTRI-Horn primary mirror were not correctly aligned.

Following this discovery, the realignment \citep{Giro} of the mirror
segments was performed by using, as feedback for the actuators, the
optical images taken with a CCD camera temporarily put on the focal
plane after removal of the ASTRI-Cherenkov camera. More conveniently,
a similar procedure could be implemented by directly using the ``variance''
information from the Cherenkov camera as feedback for the actuators
movements.

\subsection{Astrometric calibration of the camera FoV}

A second important application of the ``variance'' maps is the astrometric
calibration of the camera FoV; this is done by keeping the telescope
in staring mode (i.e. fixed azimuth. and elevation coordinates) and
acquiring ``variance'' data for a time long enough that, thanks
to the apparent rotation of the sky, numerous bright stars cross the
telescope FoV in different positions.

\begin{figure}[h]
\begin{centering}
\includegraphics[height=0.35\textwidth]{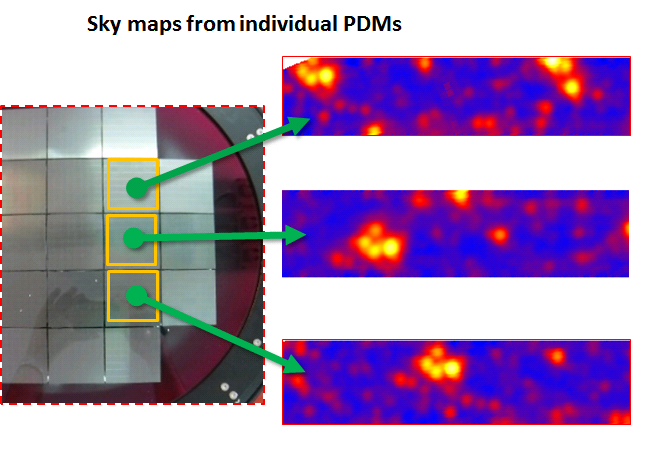}~~~~\includegraphics[height=0.35\textwidth]{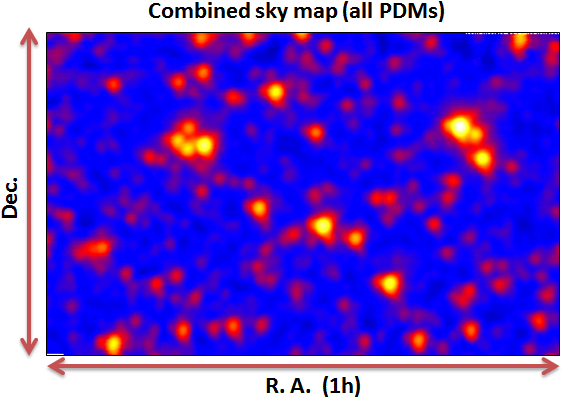}
\par\end{centering}
\caption{On the left, example of sky maps (in equatorial coordinates) generated
by using, for each one, only data relative to individual ASTRI-Horn
camera modules (PDM), acquired by keeping the telescope at fixed azimuth
and elevation coordinates; several stars are visible in multiple sky
maps since the horizontal axis is not only R.A. but also indicates
the elapsed time of the measurement. On the right panel the sky map
(Leo Minoris region) obtained by overlapping in R.A.\protect\nobreakdash-Dec.
coordinates all the PDM images after astrometric calibration of the
whole detector plane.}

\label{fig:pdm_skymaps}
\end{figure}

The sky maps obtained by mosaicking the data acquired by multiple
subsets of the camera pixels allow to determine their actual pointing
coordinates and, combining all the information, to derive the actual
camera plate scale, axes orientation and any eventual deformation
of the FoV at large off-axis angles. For example, in Fig.~\ref{fig:pdm_skymaps}
on the left panel, we show the sky maps (in equatorial R.A. - Dec.
coordinates) obtained in $\thickapprox\unit[1]{hour}$ of sky observation
from some individual PDMs ($\unit[8\mathbin{\times}8]{pixel}$) with
the telescope kept in a fixed position; the skymaps maps partially
overlap each other since, as time passes, the same sky region is observed
by multiple PDMs.

\begin{figure}[h]
\begin{centering}
\includegraphics[height=0.34\textwidth]{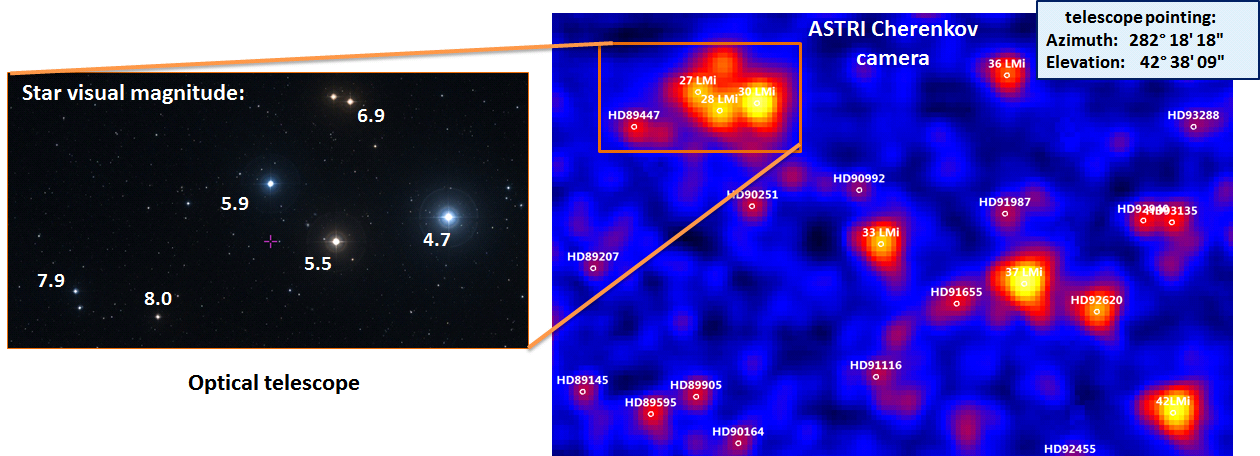}
\par\end{centering}
\caption{Right: sky map (equatorial R.A.\protect\nobreakdash-Dec. coordinates)
where the expected stars positions are shown. On the left panel a
sky map region as seen by an optical telescope were the stars visual
magnitude is indicated.}

\label{fig:Leo_Minoris_map}
\end{figure}

This procedure can be repeated with the telescope pointing in different
azimuth and altitude angles, to verify if there is any modification
of the FoV due to e.g.~static deformation of the telescope structure
caused by gravity; the results obtained can then be used to improve
the telescope pointing model.

To show the sensitivity to star flux that can be reached with the
``variance'' method, in the right panel of Fig.~\ref{fig:Leo_Minoris_map}
we show a sky map with the star identification and, on the left panel,
an optical telescope image where the star visual magnitude is indicated;
notwithstanding the quite low, $\mathrel{\thickapprox}42^{\circ}$,
elevation pointing angle in this observation (to which corresponds
an elevated atmospheric attenuation) and the reduced mirror collecting
area (as 2 mirror segments that could not be aligned were covered
with black fabric), all known stars up to $8^{th}$ visual magnitude
are clearly detected.

\subsection{Calibration and monitoring of the telescope pointing}

Another important application of the ``variance'' data is as auxiliary
tool to monitor the telescope pointing during tracking observations.
In fact, although a small CCD camera (PMC), placed on the support
structure of the ASTRI-Horn secondary mirror, provides high angular
resolution ($\unit[7.25]{arcsec}$ per pixel) images of the sky where
the telescope is pointing, its optical axis may not be precisely aligned
with the geometric center of the Cherenkov camera.

\begin{figure}[h]
\begin{centering}
\includegraphics[height=0.34\textwidth]{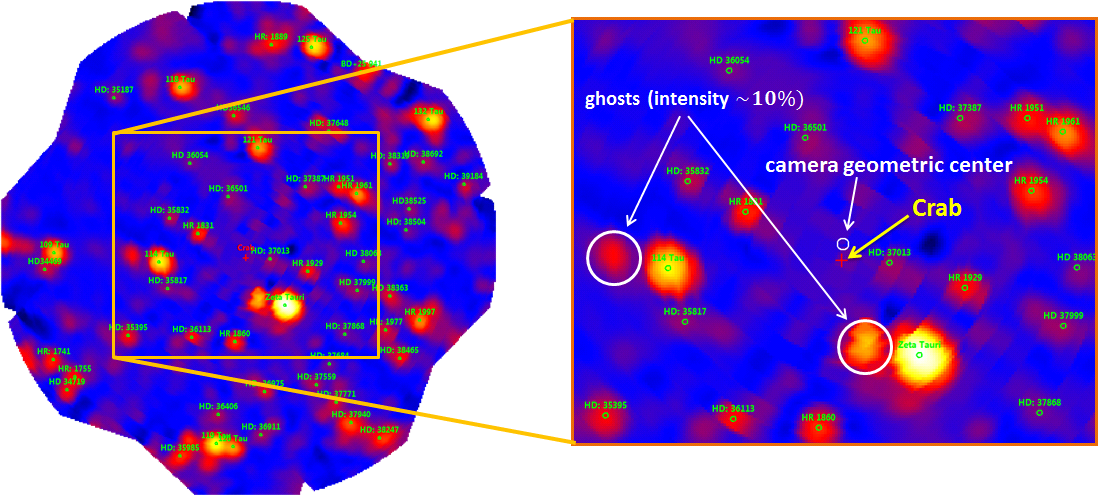}
\par\end{centering}
\caption{Left: the FoV sky map, in equatorial coordinates, obtained from the
ASTRI-Horn \textquotedblleft variance\textquotedblright{} data acquired
simultaneously to a scientific observation tracking the Crab Nebula.
Right: an enlarged view of the sky map center showing the offset between
the actual Crab position and the point correspondent to the geometric
center of the camera.}

\label{fig:Crab_Tracking}
\end{figure}

To show an example of this problem observed on real data, Fig.~\ref{fig:Crab_Tracking}
presents the sky map of the telescope FoV generated from the \textquotedblleft variance\textquotedblright{}
data acquired during a scientific observation tracking the Crab Nebula.
While the Crab optical emission is too weak to be directly observed,
the positions of the other stars in the FoV allow us to calibrate
the sky map axes and therefore to locate the actual Crab position.
As evident from the enlarged view shown in the right panel in Fig.~\ref{fig:Crab_Tracking},
in this observation a significant offset ($\thickapprox\unit[10]{'}$)
was found between the Crab position and the sky coordinates correspondent
to the geometric center of the camera (where it should have been located).
In the next section present the method used to determine the origin
of the observed pointing offset.

\subsection{Analysis of star trajectories in camera coordinates}

\begin{figure}[h]
\begin{centering}
\includegraphics[height=0.34\textwidth]{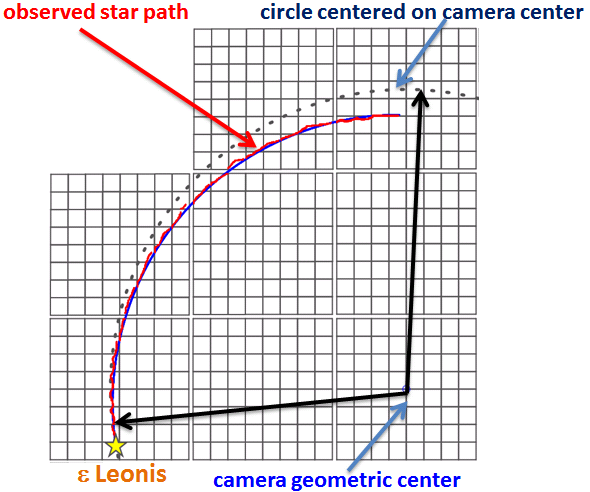}\includegraphics[height=0.34\textwidth]{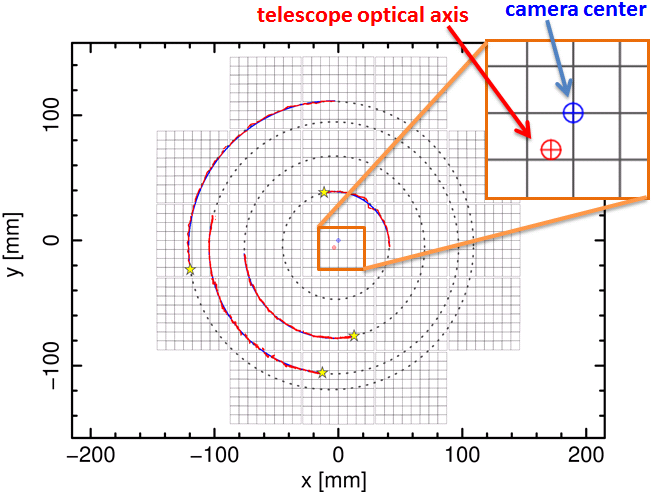}
\par\end{centering}
\caption{Left: the trajectory in camera coordinates of the star $\unit[\epsilon]{Leonis}$
as observed during a $\thickapprox\unit[4]{h}$ long tracking observation
compared to the trajectory expected if the telescope optical axis
were coincident with the camera center. Right: the path of 4 stars
(which are concentric circles) observed during the observation, from
which it is possible to determine the telescope axis offset.}

\label{fig:Crab_off_axis_Tracking}
\end{figure}

The best way to individuate the origin of systematic pointing errors
observed during tracking observations is to analyze the star trajectories
in camera coordinates that, due to the field rotation effect, should
be circles centered on the actual telescope pointing direction. For
example, in Fig.~\ref{fig:Crab_off_axis_Tracking}, we show the star
trajectories in a $\thickapprox\unit[4]{h}$ long ``Crab off-axis''
observation; in the right panel we compare the trajectory of the brightest
star in the camera FoV, $\unit[\epsilon]{Leonis}$, to the circular
trajectory that the star would have followed if the telescope pointing
had been coincident with the geometric center of the camera.

The apparent increasing drift between the observed and the expected
star positions is due to the fact that the telescope is tracking a
sky coordinate that is not coincident with the camera center. Indeed,
as shown in the left panel of Fig.~\ref{fig:Crab_off_axis_Tracking},
by the analysis of multiple star trajectories it is possible to localize
the geometrical point corresponding to the telescope tracking direction
with an accuracy that (depending on the number of stars and observing
time) can reach a few arcsec level. The measured offset can therefore
be inserted in the telescope pointing model to correct this systematic
pointing error.

\subsection{Accuracy of using stars to monitor the telescope pointing}

Once the astrometric calibration of the camera FoV has been performed,
we can use the star positions to periodically monitor the actual pointing
direction of the telescope. To reduce the systematic errors in star
localization due to the coarse pixel size of the Cherenkov camera
($\unit[11.5]{arcmin\mathbin{/}pixel}$), it is necessary to simultaneously
analyze the positions of several stars in the FoV and to average their
position over a period of time long enough that they change appreciably
their positions in camera coordinates, thanks to the field rotation.

\begin{figure}[h]
\begin{centering}
\includegraphics[height=0.34\textwidth]{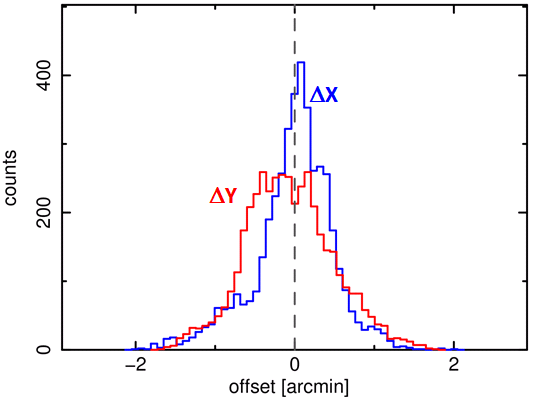}\quad{}\includegraphics[height=0.34\textwidth]{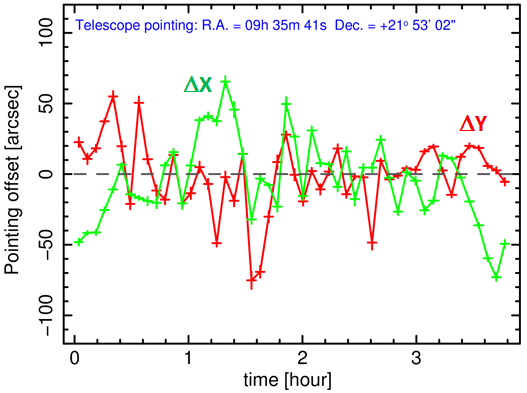}
\par\end{centering}
\caption{Left: statistical distribution of telescope pointing offsets measured
in individual variance acquisition frames. Right: the temporal evolution
of the telescope pointing offset as determined from the positions
of a few stars in the FoV, averaged over 5 minutes.}

\label{fig:tracking errors}
\end{figure}

As an example of the accuracy that can be obtained by the analysis
of star positions, in the left panel of Fig.~\ref{fig:tracking errors},
we show the histograms of the offset between the nominal telescope
pointing direction and the pointing estimated from the localization
of the 4 brightest stars in individual ``variance'' sky maps acquired
during a tracking observation. In the right panel of Fig.~\ref{fig:tracking errors}
we show the pointing offset obtained, as a function of time, by averaging
the positions of the 4 stars over a 5 minutes observing interval;
the results show how a few arcsec level accuracy can be reached in
a few minutes.

\subsection{PSF monitoring}

During tracking observations all the stars move at a very slow speed
(a few $\mu m\mathord{/}sec$) in camera coordinates, thus the ``variance''
of the pixels crossed by a star provide, with high spatial resolution,
direct information on the optical point spread function (PSF); of
course the curve observed as a function of the star position is not
directly the profile of the optical PSF, but the total star light
collected by the selected pixel. The actual shape of the optical PSF
can be however easily recovered by means of analytical models or Monte
Carlo simulation, whose parameters are tuned until a good match between
the simulated and the observed star transit profiles are obtained.

\begin{figure}[h]
\begin{centering}
\includegraphics[height=0.3\textwidth]{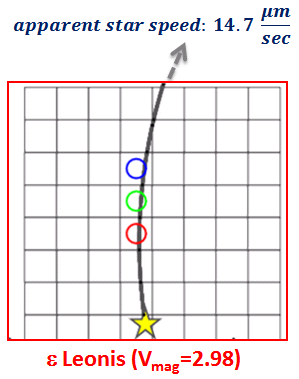}\includegraphics[height=0.3\textwidth]{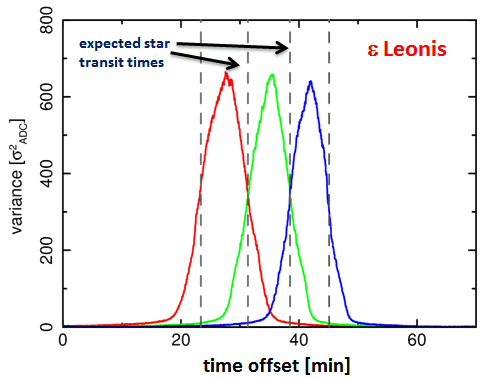}\includegraphics[height=0.3\textwidth]{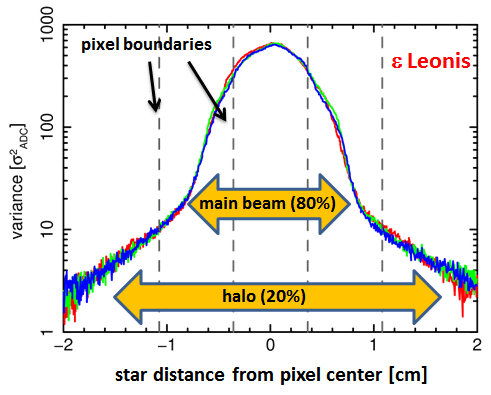}
\par\end{centering}
\caption{Left: trajectory on the focal plane of the star $\unit[\epsilon]{Leonis}$
during a tracking observation. Center: ``variance'' curves observed
during the star transit over the three pixels. Right: transit profiles
of the three pixels, as a function of the star offset from pixel centers,
showing the significant degradation of the PSF observed after a volcanic
eruption.}

\label{fig:psf}
\end{figure}

As example, in the left panel of Fig.~\ref{fig:psf} we show the
trajectory on the focal plane of a star observed during a tracking
observation and, in the central panel, the temporal evolution of the
``variance'' of three pixels located along the star trajectory.
The vertical dashed lines, indicating the expected transit times between
adjacent pixels, are computed from the telescope pointing and camera
geometry, and allow to verify the astrometric calibration of the FoV
and the telescope pointing correction performed.

On the right panel, the curves of the three pixels are shown overlapped,
in logarithmic scale, as a function of the offset of the star from
the relative pixel center. It is evident in this acquisition, the
presence on an extended halo (light scattered out of main beam) which
is due to the highly degraded mirror conditions after a volcanic eruption.
This example shows how, using the ``variance data'' it is possible
to monitor in a straightforward way the PSF with a high dynamic range
and a spatial accuracy by far superior to other methods.

\subsection*{Conclusion}

We have presented the statistical method implemented in the FPGA logic
of the ASTRI-Horn telescope that, by calculating the variance of the
ADC pedestal noise in real time, is used to obtain optical maps of
the sky in the telescope FoV with excellent accuracy and without requiring
any additional hardware.

For the very first time, our measurements with the ASTRI-Horn telescope
have demonstrated that, with a proper processing of the stars positions
as directly observed from the Cherenkov camera it is possible, notwithstanding
the coarse angular resolution of individual sky maps, to perform the
full astrometric calibration of the telescope FoV and periodically
monitor its pointing direction at a few arcsec level accuracy.

Moreover, we have shown that, by using the signal generated by the
stars during their transits over pixels, it is possible to monitor
the evolution of the optical PSF with a spatial resolution and dynamical
range by far superior to any other method.

\end{document}